# Spin polarized tunneling in the half-metallic ferromagnet $La_{0.7}Sr_{0.3}MnO_3$: experiment and theory

P. Raychaudhuri,† K. Sheshadri, P. Taneja, S. Bandyopadhyay*, P. Ayyub,
A. K. Nigam, and R. Pinto

*Tata Institute of Fundamental Research, Homi Bhabha Road, Mumbai 400005, India.*

***Abstract:*** The magnetoresistance (MR) in polycrystalline colossal magnetoresistive compounds follows a behavior different from single crystals below the ferromagnetic transition temperature. This difference is usually attributed to spin polarized tunneling at the grain boundaries of the polycrystalline sample. Here we derive a theoretical expression for the contribution of spin polarized tunneling to the magnetoresistance in ferromagnetic systems under the mean field approximation. We apply this model to our experimental data on the half metallic ferromagnet $La_{0.7}Sr_{0.3}MnO_3$, and find that the theoretical predictions agree quite well with the observed dependence of the spin polarized MR on the spontaneous magnetization.

†e-mail: prat@tifr.res.in
*Permanent Address: Indian Institute of Technology, Kharagpur, India.

# I. Introduction

The large amount of recent activity on colossal magnetoresistive (CMR) manganites has revived interest in the study of electrical transport in granular itinerant ferromagnets. Many of these materials (*e.g.*, $La_{0.7}Sr_{0.3}MnO_3$, $CrO_2$, $Fe_3O_4$) in the polycrystalline form, are now known to exhibit large magnetoresistance at low fields below the ferromagnetic transition temperature ($T_c$).[1,2,3,4]. Under similar conditions, magnetoresistance in single crystals is either very low or totally absent. The origin of this non-intrinsic intergranular effect is interesting and its detailed understanding should be important from the technological point of view.

Though several researchers have attributed this effect to intergranular spin polarized tunneling,[1,5,6,7] a complete understanding of this phenomenon is lacking. Comparing the magnetoresistance of single crystalline and polycrystalline samples of $La_{0.7}Sr_{0.3}MnO_3$, Hwang *et al.*[1] were the first to suggest that spin polarized tunneling at the polycrystalline grain boundaries might play a crucial role in determining the magnetotransport properties below the ferromagnetic $T_c$. They suggested that the magnetoresistance in polycrystalline $La_{0.7}Sr_{0.3}MnO_3$ originates from the following two sources: (i) an intrinsic part arising from Zener double exchange mechanism[8] between two neighboring manganese ions, and (ii) the intergranular spin polarized tunneling. The second component produces a sharp drop in resistance at low fields and is dominant at temperatures much below $T_c$, whereas the first component is dominant close to $T_c$. In an earlier paper[6] we have shown that the intergranular part in the magnetoresistance can be distinguished from the intrinsic part by considering the field dependence of the magnetoresistance. The contribution from the intergranular spin polarized tunneling gives rise to a sharp drop in resistance at low fields and low temperatures. At high fields, the resistance varies almost linearly with the field. The high field behavior can be attributed to the intrinsic Zener double exchange mechanism and explains the observations in single crystals. Thus one can estimate the resistance drop due to spin polarized tunneling by finding the intercept obtained by extrapolating back the linear high field region. Using this scheme, we had earlier shown that the intergranular magnetoresistance due to spin polarized tunneling in polycrystalline colossal magnetoresistive manganites drops monotonically with temperature, whereas the intrinsic part follows the behavior expected from Zener double exchange.[6]

In an attempt to understand the mechanism of magnetoresistance in granular ferromagnets, Hellman and Abeles[9] had proposed a model based on intergranular spin polarized tunneling. This model assumed that when an electron tunnels across a boundary between two grains having antiparallel magnetization, it will experience a potential barrier of the order of the exchange energy ($E_m$). So the tunneling probability between the two grains with antiparallel magnetization was assumed to be reduced by a factor $e^{-E_m/kT}$ from that when they are parallel. The conceptual difficulty in Hellman and Abeles' model is the following. When the conduction band in the ferromagnet is partially polarized, the dominant factor guiding the tunneling probability is the respective up and down density of states (DOS) in the two grains. Inoue *et al.*[10] had pointed out the difficulty in taking an exponential factor in which the energy is of the order of exchange energy. Thus, the mechanism of spin polarized transport in granular ferromagnets needs to be investigated in greater detail.

In this paper, we propose a model to understand the mechanism of spin polarized tunneling in granular ferromagnets. Section II describes the theoretical model, while the sample preparation and experimental procedures are given in Section III. In Section IV, we compare the predictions of the theoretical model with the temperature and field dependence of spin polarized tunneling in $La_{0.7}Sr_{0.3}MnO_3$. We also discuss the applicability of this model to other granular ferromagnets.

## II. Theoretical model

In a granular ferromagnet in zero field, the magnetization directions of the grains are randomly oriented due to the random orientation of their magnetocrystalline anisotropy axes. Following Inoue *et al.*,[10] we first derive an expression for the difference in the spin polarized tunneling resistance between a configuration where the magnetization of two grains are at an angle $\theta$, and that in which their magnetizations are made mutually parallel by the application of a magnetic field. Let $n_\uparrow$ and $n_\downarrow$ be the respective densities of states of up and down spins at the Fermi level. We choose our *z*-axis to be parallel to the magnetization direction of grain 1 (see Fig. 1). The up and down spin eigenstates are defined as $|S_z;\uparrow\rangle = \begin{pmatrix}1\\0\end{pmatrix}$ and $|S_z;\downarrow\rangle = \begin{pmatrix}0\\1\end{pmatrix}$ respectively. We define the spin eigenstates in grain 2 as $|S_\theta;\uparrow\rangle$ and $|S_\theta;\downarrow\rangle$. Using the Pauli rotation matrix $R(\theta)$, we can easily find the relation between $|S_\theta;\uparrow\downarrow\rangle$ and $|S_z;\uparrow\downarrow\rangle$ from

$$|S_\theta;\uparrow\downarrow\rangle = R(\theta)\,|S_z;\uparrow\downarrow\rangle,\ \text{where } R(\theta) = \begin{pmatrix}\cos(\theta/2) & -\sin(\theta/2)\\ \sin(\theta/2) & \cos(\theta/2)\end{pmatrix}. \qquad (1)$$

Hence, we obtain

$$\begin{aligned} |S_\theta;\uparrow\rangle &= \cos(\theta/2)\,|S_z;\uparrow\rangle + \sin(\theta/2)\,|S_z;\downarrow\rangle,\\ |S_\theta;\downarrow\rangle &= -\sin(\theta/2)\,|S_z;\uparrow\rangle + \cos(\theta/2)\,|S_z;\downarrow\rangle. \end{aligned} \qquad (2)$$

An electron can tunnel from grain 1 to grain 2 using one of the following channels:

$$\text{(a)} \qquad |S_z;\uparrow\rangle \rightarrow |S_\theta;\uparrow\rangle$$

$$\text{(b)} \qquad |S_z;\uparrow\rangle \rightarrow |S_\theta;\downarrow\rangle$$

$$\text{(c)} \qquad |S_z;\downarrow\rangle \rightarrow |S_\theta;\uparrow\rangle$$

$$\text{(d)} \qquad |S_z;\downarrow\rangle \rightarrow |S_\theta;\downarrow\rangle.$$

Since the Hamiltonian involved with the tunneling process is spin independent, the matrix elements corresponding to (a) and (d) should be proportional to $\cos^2(\theta/2)$, and the matrix element corresponding to (b) and (c) should be proportional to $\sin^2(\theta/2)$. The total transition probability of up (or down) spin for tunneling from grain 1 to grain 2 should also depend on the initial and final DOS available to it in the two grains. Thus, the total transition probabilities for the four processes will be

$$T_a \propto n_\uparrow^2\cos^2(\theta/2)$$

$$T_b \propto n_\uparrow\, n_\downarrow\sin^2(\theta/2)$$

$$T_c \propto n_\uparrow\, n_\downarrow\sin^2(\theta/2)$$

$$T_d \propto n_\downarrow^2\cos^2(\theta/2).$$

The tunneling conductivity, $\sigma(\theta)$, between the two grains at angle $\theta$ involve the sum of all four processes. Thus,

$$\sigma(\theta) \propto (n_\uparrow^2+n_\downarrow^2)\cos^2(\theta/2) + 2n_\uparrow\, n_\downarrow\sin^2(\theta/2)$$

$$\text{or,} \qquad \sigma(\theta) \propto (1/2)(n_\uparrow+n_\downarrow)^2[1+P^2\cos\theta], \qquad (3)$$

where $P = (n_\uparrow - n_\downarrow)/(n_\uparrow+n_\downarrow)$. When the magnetizations of grain 1 and grain 2 are made parallel to each other by applying a magnetic field, the conductivity becomes:

$$\sigma(\theta=0) \propto (n_\uparrow^2+n_\downarrow^2). \qquad (4)$$

In a real system with many grains and grain boundaries one has to average over cosθ. Thus, the resistance change ($\Delta R_{spt}$) arising due to spin polarized tunneling is proportional to [1/σ(θ)–1/σ(0)].

In order to determine the temperature dependence of $\Delta R_{spt}$, we require the temperature dependence of $n_\uparrow$ and $n_\downarrow$. We use the ferromagnetic Kondo Hamiltonian initially proposed for this system by Furukawa[11,12,13] to calculate the up and down DOS.

$$H = -t \sum_{\langle i,j \rangle} c^\dagger_{i\sigma} c_{j\sigma} - J_H \sum_i \boldsymbol{\sigma}_i \cdot \mathbf{s}_i$$

where $t$ is the nearest neighbor hopping energy of the $e_g$ electrons and $J_H$ is the local ferromagnetic Hund's rule coupling between the $e_g$ electron spin $\boldsymbol{s}_i$ and the $t_{2g}$ spin $\boldsymbol{\sigma}_i$ at the i-th site. In terms of $\boldsymbol{\sigma}^{\pm} = \sigma^x \pm i\sigma^y$, $\sigma^z$, $\mathbf{s}^{\pm} = s^x \pm i s^y$ we can write,

$$H = -t \sum_{\langle i,j \rangle} c^\dagger_{i\sigma} c_{j\sigma} - \frac{J_H}{2} \sum_i (\sigma_i^+ s_i^- + \sigma_i^+ s_i^- + 2 \;\; \sigma_i^z s_i^z)$$

We resort to a mean field approximation in which the $t_{2g}$ spin $\boldsymbol{\sigma}$ is treated as a classical vector, i.e., the operators $\sigma^{\pm}$ and $\sigma^z$ are replaced by c-numbers:

$$\sigma_i^- \equiv A \;, \quad \sigma_i^+ \equiv A^* , \text{ and } \sigma_i^z \equiv B \;.$$

We also write the $e_g$ spin operators in terms of annihilation and creation operators:

$$s_i^+ = c^\dagger_{i\uparrow} c_{i\downarrow} , \; s_i^- = c^\dagger_{i\downarrow} c_{i\uparrow} \text{ and } s_i^z = \frac{1}{2}(c^\dagger_{i\uparrow} c_{i\uparrow} - c^\dagger_{i\downarrow} c_{i\downarrow}) .$$

As a result, we get the following expression for the mean field Hamiltonian:

$$H^{MF} = -t \sum_{\langle i,j \rangle} c^\dagger_{i\sigma} c_{j\sigma} - \frac{J_H}{2} \sum_i [A\, c^\dagger_{i\uparrow} c_{i\downarrow} + A^* c^\dagger_{i\downarrow} c_{i\uparrow} + B(c^\dagger_{i\uparrow} c_{i\uparrow} - c^\dagger_{i\downarrow} c_{i\downarrow})] .$$

Going over to ***k***-space by Fourier transforming according to

$$c_{i\sigma}^{\dagger} = \sum_{\boldsymbol{k}} c_{\boldsymbol{k}\sigma}^{\dagger} e^{i\boldsymbol{k}.\boldsymbol{r}}$$

we get,

$$H^{MF} = \sum_{k}\left[\left(-\gamma_{\boldsymbol{k}} - \frac{1}{2}\mathrm{B}\,J_H\right)c_{\boldsymbol{k}\uparrow}^{\dagger}c_{\boldsymbol{k}\uparrow} + \left(-\gamma_{\boldsymbol{k}} + \frac{1}{2}\mathrm{B}\,J_H\right)c_{\boldsymbol{k}\downarrow}^{\dagger}c_{\boldsymbol{k}\downarrow} - \frac{J_H}{2}\left(\mathrm{A}\,c_{\boldsymbol{k}\uparrow}^{\dagger}c_{\boldsymbol{k}\downarrow} + \mathrm{A}^{*}c_{\boldsymbol{k}\downarrow}^{\dagger}c_{\boldsymbol{k}\uparrow}\right)\right]$$

Here $\gamma_{\boldsymbol{k}} = t\sum_{\Delta} e^{i\boldsymbol{k}.\Delta}$ where Δ is the nearest neighbor vector. We can rewrite this as

$$H^{MF} = \sum_{\boldsymbol{k}}\begin{pmatrix} c_{\boldsymbol{k}\uparrow}^{\dagger} & c_{\boldsymbol{k}\downarrow}^{\dagger}\end{pmatrix} h_{\boldsymbol{k}} \begin{pmatrix} c_{\boldsymbol{k}\uparrow} \\ c_{\boldsymbol{k}\downarrow}\end{pmatrix},$$

where
$$h_{\boldsymbol{k}} = \begin{pmatrix} -\gamma_{\boldsymbol{k}} - \frac{1}{2}\mathrm{B}\,J_H & -\frac{1}{2}\mathrm{A}\,J_H \\ -\frac{1}{2}\mathrm{A}^{*}J_H & -\gamma_{\boldsymbol{k}} + \frac{1}{2}\mathrm{B}\,J_H \end{pmatrix}.$$

This 2×2 matrix is easily diagonalized to get the eigenvalues as

$$\lambda_{\boldsymbol{k}}^{\pm} = \frac{1}{2}\left(-2\gamma_{\boldsymbol{k}} \pm J_H\sigma\right),$$

where $\sigma^2 = \sigma_x^2 + \sigma_y^2 + \sigma_z^2 \equiv |\mathrm{A}|^2 + \mathrm{B}^2$. The corresponding eigenvectors are:

$$\left|u_{\boldsymbol{k}}^{-}\right\rangle = x_{\boldsymbol{k}}\left|\uparrow\right\rangle + y_{\boldsymbol{k}}\left|\downarrow\right\rangle \text{ and } \left|u_{\boldsymbol{k}}^{+}\right\rangle = y_{\boldsymbol{k}}\left|\uparrow\right\rangle - x_{\boldsymbol{k}}\left|\downarrow\right\rangle,$$

where $x_{\boldsymbol{k}}^2 = \frac{1}{2}\left(1 + \frac{\mathrm{B}}{\sigma}\right)$ and $y_{\boldsymbol{k}}^2 = \frac{1}{2}\left(1 - \frac{\mathrm{B}}{\sigma}\right)$, and, $\left|\uparrow\right\rangle = \begin{pmatrix}1\\0\end{pmatrix}$ and $\left|\downarrow\right\rangle = \begin{pmatrix}0\\1\end{pmatrix}$.

At zero temperature the $t_{2g}$ spin assumes a saturation value, so $B/\sigma = 1$, and as a result the ground state $\left|u_k^{-}\right\rangle$ has purely up-spin character. The density of states corresponding to T=0 is shown schematically in Fig. 2(a). For T > 0, $B/\sigma < 1$ and as a result the ground state and excited state have mixed character:

$$N_{\uparrow}(E) = x_k^2\ N(E), \text{ and } N_{\downarrow}(E) = y_k^2\ N(E).$$

The up and down spin characters, $x_{\mathbf{k}}^2$ and $y_{\mathbf{k}}^2$, respectively, are both non-zero in this case, as shown in Fig. 2(b). Finally, for $T \geq T_c$, $B$ vanishes, resulting in $x_{\mathbf{k}}^2 = y_{\mathbf{k}}^2$. This leads to $N(E)$ being an equal admixture of up and down spins[14] as shown schematically in Fig. 2(c). In our theory, the magnetization of the $e_g$ spin is:

$$m_{e_g} = \frac{1}{2}\left(x_{\mathbf{k}}^2 - y_{\mathbf{k}}^2\right) = \frac{B}{2\sigma}.$$

Therefore the total (spontaneous) magnetization is $M = (B/2\sigma) + B = \alpha B$, where $\alpha = (1 + 1/2\sigma)$. We can thus write $x_{\mathbf{k}}^2$ and $y_{\mathbf{k}}^2$ in terms of the reduced magnetization, $m = M(T)/M(T \rightarrow 0)$ as:

$$x_{\mathbf{k}}^2 = (1 + m)/2, \text{ and } y_{\mathbf{k}}^2 = (1 - m)/2 \qquad (5)$$

Within our theory, the ratio $N_{\uparrow}(E)/N_{\downarrow}(E)$ is independent of energy. Thus, the up and down DOS at the Fermi level, $n_{\uparrow}$ and $n_{\downarrow}$, which appear in equations (3) and (4) are proportional to $x_{\mathbf{k}}^2$ and $y_{\mathbf{k}}^2$ respectively. Substituting equation (5) in the expression for spin polarized tunneling conductivity in equation (3) and (4), we obtain a relation between $\Delta R_{spt}$ and the reduced spontaneous magnetization of the system:

$$\Delta R_{spt} \propto [\langle 1/\sigma(\theta)\rangle - (1/\sigma(0)] \qquad (6)$$

Here the conductivity in zero field, $1/\sigma(\theta)$, is averaged over different values of $\cos\theta$ for various grains. While fitting $\Delta R_{spt}$ vs. $m$ data (see Section IV) we have used $\langle\cos(\theta)\rangle$ as a fitting parameter.

## III. Experiment

Bulk samples of $La_{0.7}Sr_{0.3}MnO_3$ were prepared by a wet chemical route. Initially, a stoichiometric mixture of $La_2O_3$ and Mn was dissolved in nitric acid and $Sr(NO_3)_2$ added to it. The cations were co-precipitated as carbonates by adding excess ammonium carbonate to the nitrate solution. This carbonate precursor was then heated at 1200°C to obtain single phase $La_{0.7}Sr_{0.3}MnO_3$. Electrical measurements were carried out on a rectangular bar shaped sample (2mm × 3mm × 15mm) using the conventional four probe technique. Magnetoresistance measurements were made using a home made superconducting magnet up to fields of 3 Tesla, in the temperature range 5 K to 300 K. Magnetization versus field isotherms were measured on a Quantum Design SQUID magnetometer up to a field of 5.5 Tesla.

## IV. Results and Discussion

Figure 3(a) shows the magnetoresistance, MR ≡ $[R(H)–R(0))/R(0]$, as a function of the applied field ($H$) from 5 K to 300 K. We observe a sharp drop in the MR at low fields and the magnitude of the drop decreases with increasing temperature. This sharp drop is absent in single crystals and epitaxial thin films[1,2,6] and is therefore associated with the grain boundaries in polycrystalline samples. At high field, the slope of this curve becomes almost linear. The slope increases with increasing temperature and is identical to the nature of the MR vs. $H$ curves observed in single crystals. This part of the MR can be ascribed to the Zener double exchange and arises from the intrinsic magnetoresistance of the grains.[6] To find the total resistance drop associated with the intergranular part ($\Delta R_{spt}$) we extrapolate back the linear high field region to find its intercept at zero field. This is schematically shown in Fig 3(b) for the curve obtained at 5 K. Figure 3(c) shows the magnetization versus field ($M$-$H$) data for the same sample from 5 K to 340 K. The inset shows the spontaneous magnetization ($M_s$) as a function of temperature calculated by back extrapolating the linear region of the curve beyond technical saturation.

In Fig. 4 we have plotted $\Delta R_{spt}$ as a function of $m$ ( = $M_s(T)/M_s(5K)$ ) along with the theoretical curve (Eq. 6). The proportionality constant in Eq. (6) and $\langle\cos\theta\rangle$ are taken as fitting parameters. We get the best fit for $\langle\cos\theta\rangle = -0.8788$, which corresponds to $\theta\approx151.5^\circ$. The negative value of $\langle\cos\theta\rangle$ is consistent with the fact that there are two interactions governing the direction of magnetization of a grain: the easy axis which is random and tends to randomize the direction of the magnetization, and the magnetostatic dipolar interaction between the grains which tend to favour antiparallel alignment between grains.

So far, we have been concerned with the total MR arising from spin polarized tunneling, that is the total drop in resistance due to spin polarized tunneling after the technical saturation in magnetization. We now investigate the MR versus $M$ at low fields. This is the regime in which domain rotation takes place, i.e. where $\langle\cos(\theta)\rangle$ is changing. The MR versus ($M/M_s$) curves at fields less than 1 kOe are shown in Fig. 6. These curves fit well with a power law behavior: MR $\propto$ $(M/M_s)^n$, where n<1. However, to fit these curves within our model, we need to compute $\langle\cos\theta\rangle$ as a function of field, and this is beyond the scope of the present work.

We have thus shown that the model of spin polarized tunneling proposed in this paper fits well with the data on the half metallic ferromagnet $La_{0.7}Sr_{0.3}MnO_3$ (with the underlying $t_{2g}$ spins forming a localized magnetic lattice and the $e_g$ spins forming a polarized conduction band). However, equations (3) and (4) should apply equally well to other granular ferromagnetic systems, provided the evolution of the up and down spin density of states with temperature is known. It might therefore be also interesting to investigate the nature of spin polarized tunneling in more conventional granular itinerant ferromagnets within the realms of this model.

## V. Conclusion

In summary, we have developed a mean field model for spin polarized tunneling in granular itinerant ferromagnets. The theoretical model provides a satisfactory fit for our data on the temperature dependence of the magnetoresistance due to spin polarized tunneling in the half metallic ferromagnet $La_{0.7}Sr_{0.3}MnO_3$. Further investigations on other systems should provide a better understanding of the transport mechanism in itinerant magnetic systems.

## Acknowledgments

We would like to thank K.V. Gopalakrishnan for his help regarding SQUID measurements and Arun Paramekanti for useful discussions. We would also like to thank R. S. Sannabhadti and Arun Patade for technical help.

## FIGURE CAPTIONS

FIGURE 1. Two neighboring ferromagnetic grains with magnetization $\mathbf{M_1}$ and $\mathbf{M_2}$ making an angle θ with respect to each other. In a sufficiently large applied field, θ becomes zero, thus increasing the tunneling probability of an electron from grain 1 to grain 2. In our convention, the z-axis is along the magnetization direction of grain 1.

FIGURE 2. A schematic diagram of the up and down densities of states at various temperatures: (a) T=0, (b) $0<T<T_c$ and (c) $T \geq T_c$. $E_F$ shows the Fermi energy for the hole doped compound. At finite temperature, the spin eigenstates have a mixed character.

FIGURE 3. (a) Magnetoresistance versus field isotherms for polycrystalline $La_{0.7}Sr_{0.3}MnO_3$ from 5 K to 300 K. The sharp drop at low fields arises from spin polarized tunnelling at the polycrystalline grain boundaries. (b) The total resistance drop due to spin polarized tunnelling ($\Delta R_{spt}$) is estimated by back extrapolating the linear high field slope and finding its intercept at zero field. (c) The magnetization versus field isotherms at different temperatures. Inset shows the temperature variation of spontaneous magnetization ($M_s$) calculated by back extrapolating linear portion curve after technical saturation. At 340 K the spontaneous magnetization is calculated from the linear portion of the Arott plot (M/H versus $M^2$).

FIGURE 4. The total resistance drop due to spin polarized tunneling ($\Delta R_{spt}$) at different temperatures as a function of the reduced magnetization $m = M_s(T)/M_s(5K)$ (filled circles). The solid line is the theoretical best fit curve (Eq. (6)) to the experimental points. The best fit value of $\langle\cos(\theta)\rangle$ is –0.8788.

FIGURE 5. Magnetoresistance versus magnetization isotherms at low fields (up to 1 kOe). The solid lines are the best fit to the relation MR $\propto (M/M_s)^n$. Note that this relation provides a good fit to the data over the entire temperature range.

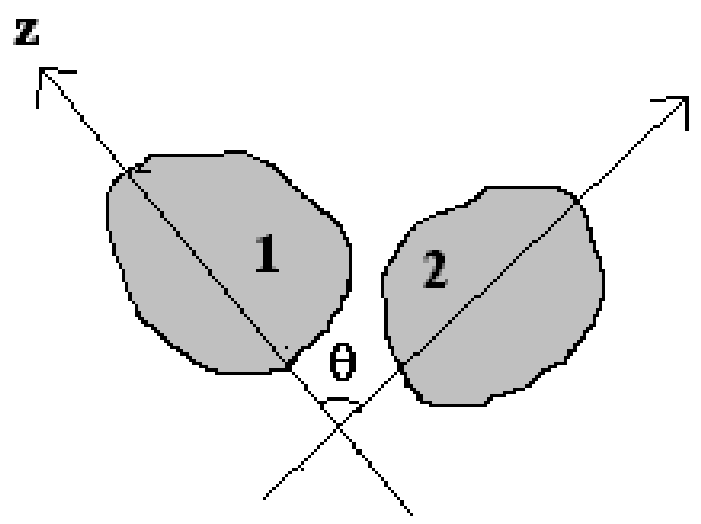

**Figure 1 (P Raychaudhuri)**

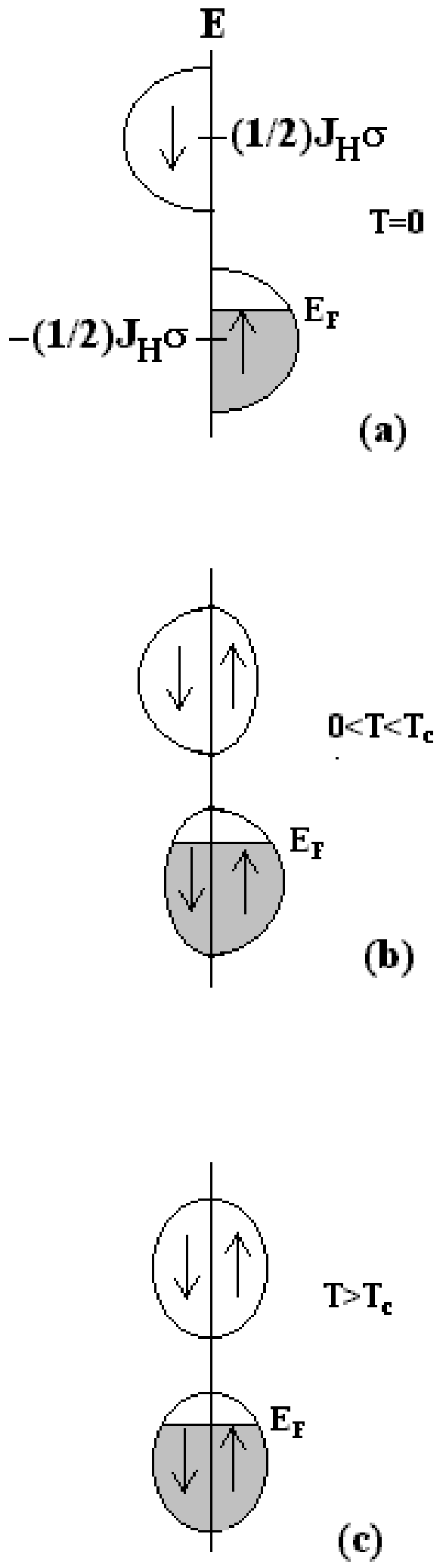

Figure 2 (P Raychaudhuri et al.)

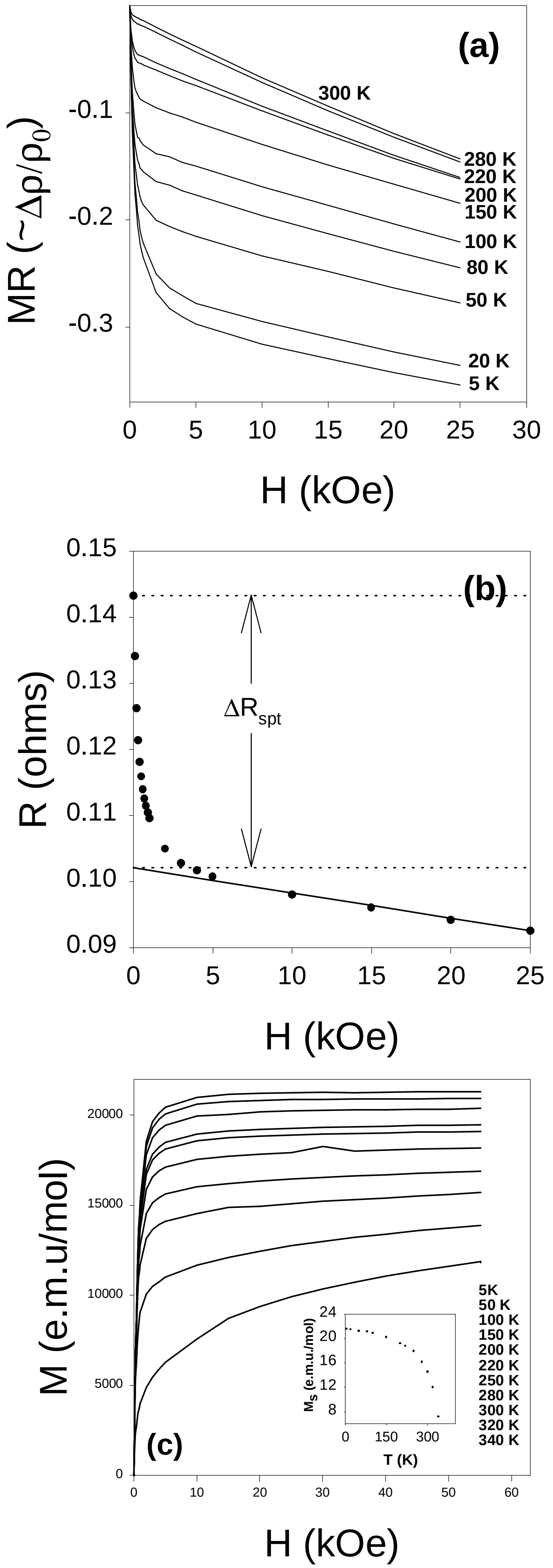

**Figure 3 (P. Raychaudhuri)**

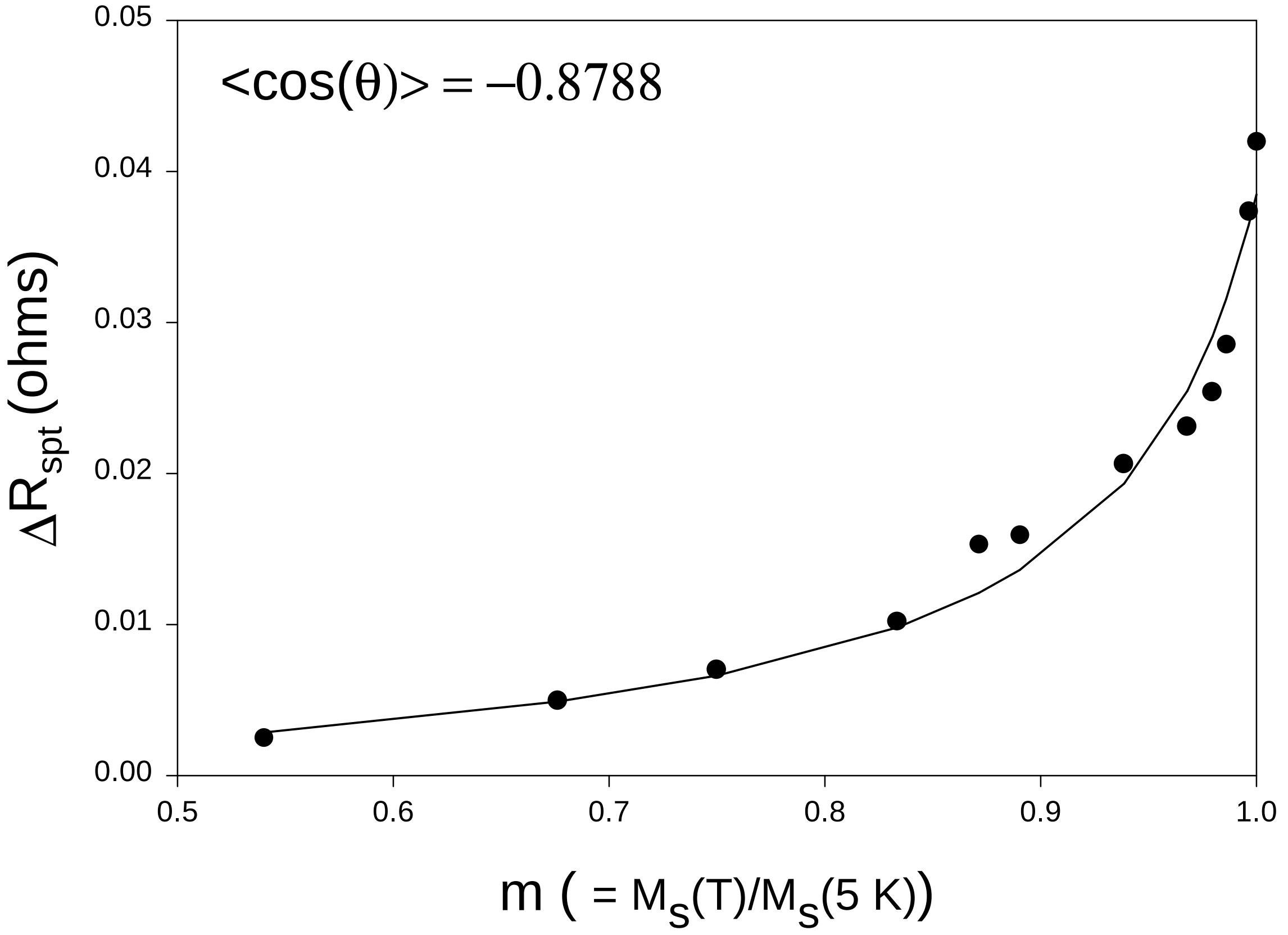

**Figure 4 (P. Raychaudhuri et al)**

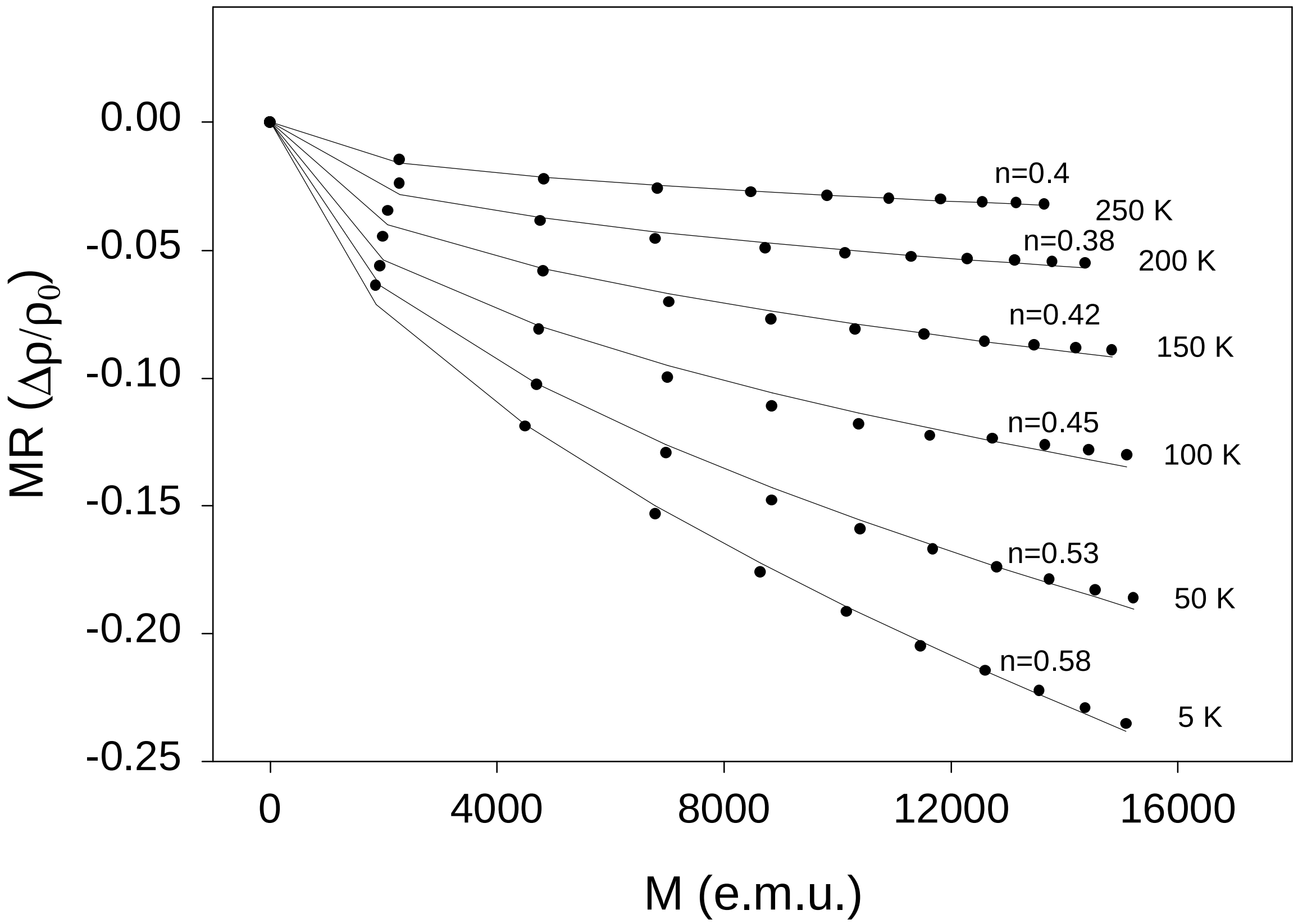

Figure 5 (P Raychaudhuri et al)